# Hidden force floating ice

Chang Q Sun, Ecqsun@ntu.edu.sg

Nanyang Technological University, Singapore

**Because of the segmental specific-heat disparity of the hydrogen bond (O:H-O) and the Coulomb repulsion between oxygen ions, cooling elongates the O:H-O bond at freezing by stretching its containing angle and shortening the H-O bond with an association of larger O:H elongation, which makes ice less dense than water, allowing it to float.**

Ref:

[1] *Mpemba effect*, http://arxiv.org/abs/1501.00765
[2] *Hydrogen-bond relaxation dynamics: resolving mysteries of water ice.* Coord. Chem. Rev., 2015. **285**: 109-165.

## 1.1 Anomaly: floating of ice

Observations shown in Figure 1 confirmed the following:

1) Ice cube is less dense than water so it floats in water [1].
2) The mass density $\rho(T)$ profile oscillates over the full temperature range [2].
3) Cooling densification proceeds in the liquid (I) and the solid (III) phase and cooling expansion occurs to the quasi-solid phase (II) and ice in the very-low temperature regime (IV) at different rates. Density transits at 277 K (for bulk), 202 K and 50 K (for droplet of 1.4 nm size).

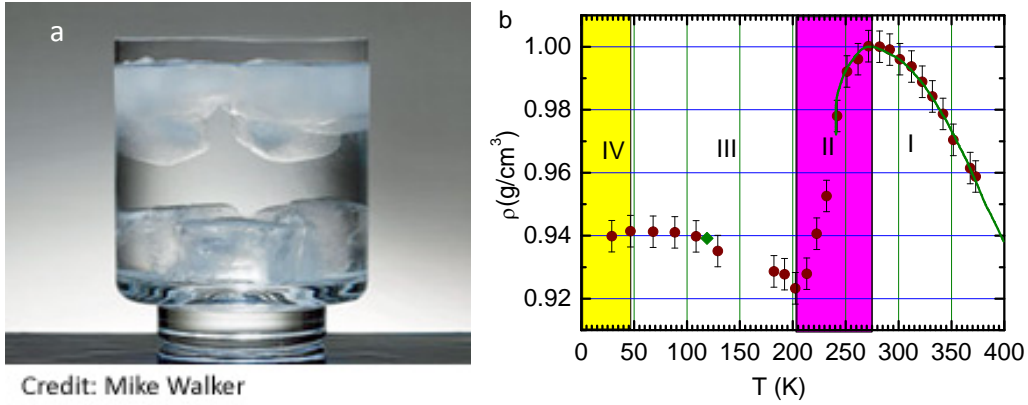

Figure 1. (a) Low density ice cubes float in a cup of water [1] and (b) the density $\rho(T)$ profile of water oscillates over the full temperature range for 1.4 nm droplet obtained using Raman and FTIR [2].

## 1.2  Reasons: O:H-O bond specific disparity

The following rationalizes the anomaly, see Figure 2 [3]:

1) A supposition of the specific heat $\eta_x(T, \Theta_{Dx})$[1] curves for the O:H nonbond (subscript x = L) and the H-O bond (x = H) defines two intersecting temperatures that divide the full temperature range into four regimes with different $\eta_L/\eta_H$ ratios, see Figure 2a.

2) Cooling stretches the ∠O:H-O containing angle θ (Figure 2b) and elongates the O-O distance in the II, III, IV regimes. The θ remains constant in liquid phase I because of the high fluctuation in molecular motion, rotation, and vibration.

3) Cooling shortens the segment with relatively lower $\eta_x(T)$ value and lengthens the other cooperatively through O-O Coulomb repulsion. The O:H contracts always more than H-O expands in opposite direction in phase I and III ($\eta_L/\eta_H < 1$). Both segments remain unchanged in IV because $\eta_L \approx \eta_H \approx 0$. In the transition regime II ($\eta_L/\eta_H > 1$), H-O contracts less than O:H elongates. The $d_x(T)$ profiles in Figure 2c is independent of θ and $\eta_L/\eta_H = 0$ contribution.

4) Converted from measurement in Figure 1b, the $d_x(T)$ profile in Figure 2d represents the true situation

---

[1] The segmental specific heat $\eta_x$ is characterized by its Debye temperature $\Theta_{Dx}$ and its integration from 0 K to the $T_{mx}$. The $\Theta_{Ds}$ approximates the saturation temperature and the integration the cohesive energy $E_x$. For the O:H-O bond, $\Theta_{DL}/\Theta_{DH} = 198/3200 = \omega_L/\omega_D = 200/3200$; $E_L/E_H = 0.1/4.0$.

being inclusive of θ relaxation and $\eta_L/\eta_H = 0$ contribution in regime IV.

5) The O:H-O bond thermodynamic relaxation in the angle and the segmental length makes ice less dense than water, allowing it to float.

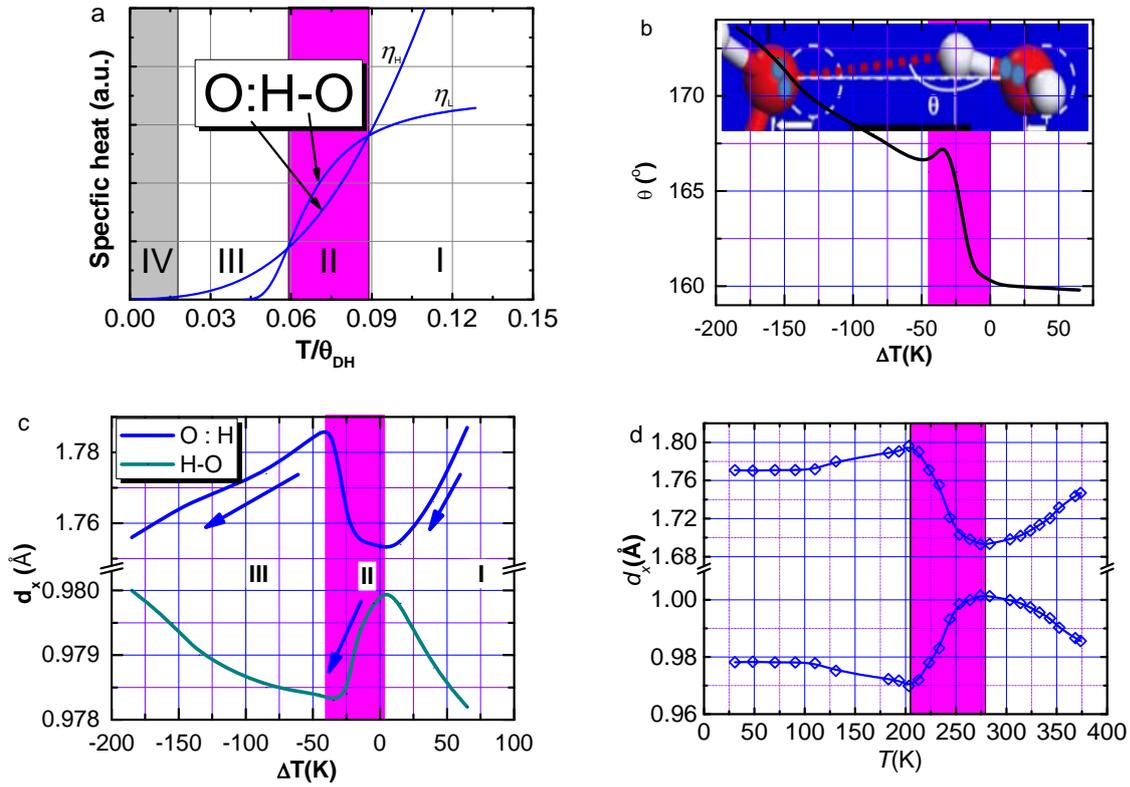

Figure 2  The superposition of the specific heats defines (a) two intersecting points that define four regimes of different $\eta_L/\eta_H$ ratios over the full temperature range [3]. (b) Cooling stretches the angle θ in different regimes at different slopes. The segment of lower $\eta_x$ serves as the master to follow the general rule of cooling contraction and the other slave part relaxes in the same direction by different amount. The O:H and the H-O relax (c) without (numerical solution) [3] and (d) with θ contribution [4]. Inset (b) illustrates the segmental cooling relaxation in the quasi-solid phase II resulting in net O:H-O length gain, which makes ice float. (Reprinted with permission from [3].)

## 1.3    Indication

### 1.3.1 Life under and above ice

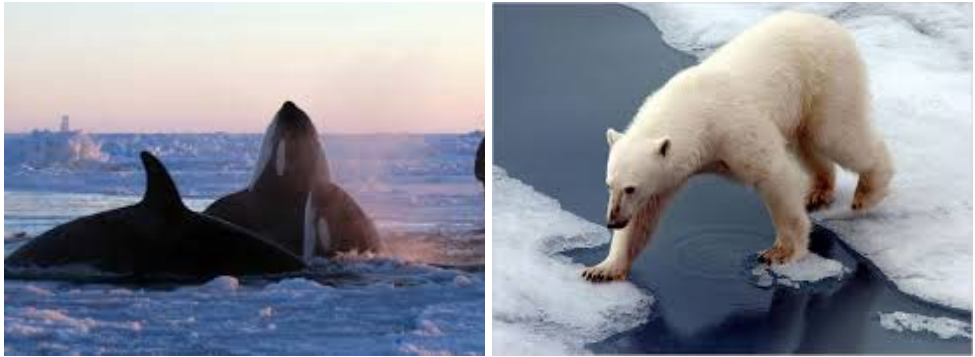

Figure 3 Floating enables surviving and reproducing of creatures in winter such whale (Credit to John Eggers, Bemidji, USA) and bears. Polar bears like this one are excellent swimmers but use floating sea ice as pathways to coastal areas and as platforms from which to hunt seals (Credited to Thomas Nilsen, The New York Times 2006).

### 1.3.2 Rock erosion: freezing-melting cycle

As noted by the Chines sage Lao Tzu in his ancient text: *"There is nothing softer and flexible than water, and yet there is nothing better for attacking hard and strong stuff."* Erosion of rocks is the nature phenomenon as shown in Figure 4 morphologies of the fresh and the eroded rock blocks. Rainfall water penetrated into the rock through pores become ice at freezing in the Autumn and Winter [5]. Volume expansion of ice enlarges the pore size, which exerts force nearby breaking the rocks. Melting of ice in the spring and evaporation of the molten ice in the Spring and Summer leaves the damage behind. Repeated occurrence erodes the rocks.

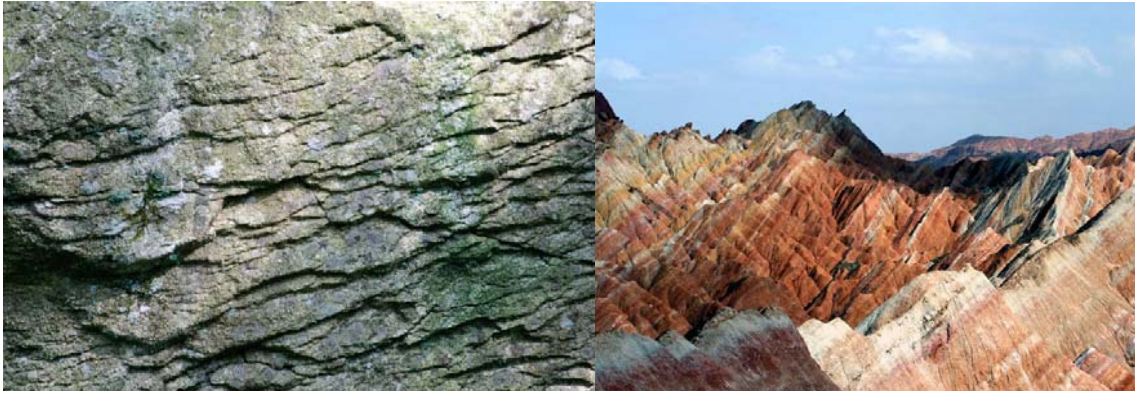

Figure 4 Photographs of eroded rocks (National geographic park, Zhangyie, Gansu, China).

Erosion is the act in which earth is worn away. A similar process, weathering, breaks down
or dissolves rock, weakening it or turning it into tiny fragments. No rock is hard enough to resist the forces of erosion. Together, they shaped the sharp peaks of the Himalaya Mountains in Asia and sculpted the spectacular forest of rock towers of Bryce Canyon, in the U.S. state of Utah, as well as Zhang Ye in Gansu, China.

The process of erosion moves bits of rock or soil from one place to another. Most erosion is performed by water, wind, or ice. These forces carry the rocks and soil from the places where they were weathered. When wind or water slows down, or ice melts, sediment is deposited in a new location. As the sediment builds up, it creates fertile land. River deltas are made almost entirely of sediment. Delta sediment is eroded from the banks and bed of the river.

### 1.3.3  Watering soil in winter –freezing expansion

Watering soil in winter has many advantages for keeping nutrition and fertilizing the soil [6]. Water molecules of snows or watering penetrate into the earth and freeze in cold weather. Water freezing expands its volume and the soil. Ice melting and evaporating leave pores to loosen and soften the solid with reservation of nutrition and moisture, which is beneficial for the plant growth in the next Spring.

### 1.3.4  Sea level rise – global warming

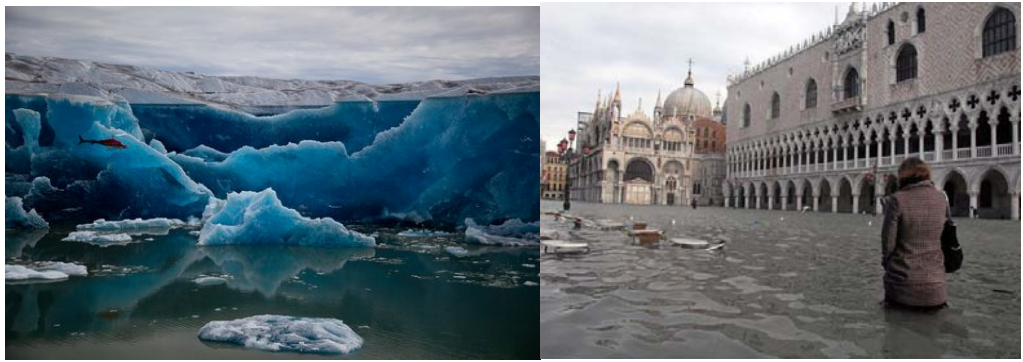

Figure 5 (a) Ice melting due to Global warming (left) raises the sea levels worldwide at a rate of 3.5 mm per year since the early 1990s. The trend puts thousands of coastal cities, like Venice (right), Italy, (seen here during a historic flood in 2008), and even whole Greenland islands at risk of being claimed by the ocean [7].

Every degree Fahrenheit of global warming due to carbon pollution, global average sea level will rise by about 4.2 feet in the long run by ice melting [8]. When multiplied by the current rate of carbon emissions, and the best estimate of global temperature sensitivity to pollution, this translates to a long-term sea level rise commitment that is now growing at about 1 foot per decade. Such rates, if sustained, would realize the highest levels of sea level rise contemplated here in hundreds, not thousands of years — fast enough to apply continual pressure, as well as threaten the heritage, and very existence, of coastal communities everywhere.

Over the past century, the burning of fossil fuels and other human and natural activities has released enormous amounts of heat-trapping gases into the atmosphere. These emissions have caused the Earth's surface temperature to rise, and the oceans absorb about 80 percent of this additional heat. The rise in sea levels is linked to three primary factors, all induced by this ongoing global climate change:

1) **Thermal expansion:** When water heats up, it expands. About half of the past century's rise in sea level is attributable to warmer oceans simply occupying more space.
2) **Melting of glaciers and polar ice caps:** Large ice formations, like glaciers and the polar ice caps, naturally melt back a bit each summer. But in the winter, snows, made primarily from evaporated seawater, are generally sufficient to balance out the melting. However, persistently higher temperatures caused by global warming have led to greater-than-average summer melting as well as diminished snowfall due to later winters and earlier springs. This imbalance results in a significant net gain in runoff versus evaporation for the ocean, causing sea levels to rise.

3) **Ice loss from Greenland and West Antarctica:** As with glaciers and the ice caps, increased heat is causing the massive ice sheets that cover Greenland and Antarctica to melt at an accelerated pace. Meltwater from above and seawater from below is seeping beneath Greenland's and West Antarctica's ice sheets, effectively lubricating ice streams and causing them to move more quickly into the sea. Moreover, higher sea temperatures are causing the massive ice shelves that extend out from Antarctica to melt from below, weaken, and break off.

When sea levels rise rapidly, as they have been doing, even a small increase can have devastating effects on coastal habitats. As seawater reaches farther inland, it can cause destructive erosion, flooding of wetlands, contamination of aquifers and agricultural soils, and lost habitat for fish, birds, and plants. When large storms hit land, higher sea levels mean bigger, more powerful storm surges that can strip away everything in their path. In addition, hundreds of millions of people live in areas that will become increasingly vulnerable to flooding. Higher sea levels would force them to abandon their homes and relocate. Low-lying islands could be submerged completely.

## 1.3  History

Ice floating follows Archimedes' principle which indicates that the upward buoyant force (B) that is exerted on a body immersed in a fluid, whether fully or partially submerged (V+ΔV), is equal to the weight of the fluid that the body displaces $Vg\rho_{liquid}$. The ΔV is the unsubmerged volume. The following formulates the net force $f$ floating the body:

$$f = B - Mg = \left[V\rho_{liquid} - (V+\Delta V)\rho_{body}\right]g$$
$$= V\rho_{liquid}g\left[1 - \frac{(V+\Delta V)\rho_{body}}{V\rho_{liquid}}\right] \geq 0$$

Which requires,

$$\frac{\rho_{liquid} - \rho_{body}}{\rho_{body}} = \frac{\Delta\rho}{\rho_{body}} \geq \frac{\Delta V}{V}.$$

That is, the density of the body is smaller than the that of the liquid. Therefore, ice is less dense than water, allowing it to float, as everybody knows.

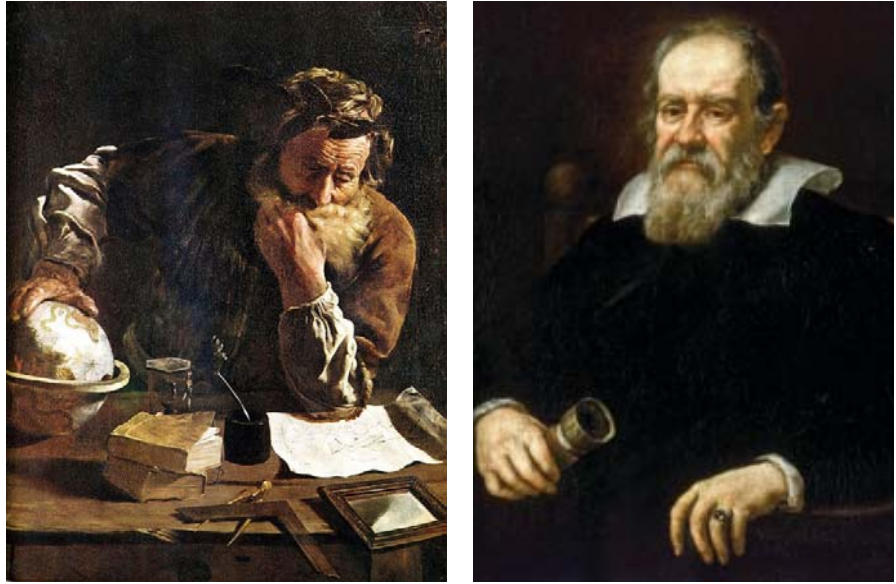

Figure 6 Painting portrait of Archimedes by Domenico Fetti in 1620 and Galili Galileo by Justus Susterman in 1636 (from free Wikimedia). Archimedes (Syracuse, Italy. 287-212 B.C.) was a greek mathematician, physicist, engineer, inventor, and astronomer. Galili Galileo (Pisa, Italy. 1564 – 1642) was a physicist, mathematician, engineer, astronomer, and philosopher.

In 1611[9], Galileo and Ludovico delle Colombe had a fierce, multiday debate on the topic "Why does ice float on water, when ice is itself water?" in front of dozens of wealthy spectators gathered in Florence. Ludovico (Florence, Italy, 1565 – 1616) was a philosopher and a poet. Ludovico is known above all for his opposition to Galileo, at first in the field of astronomy in his siding against the Copernican system (Earth revolved around the Sun), and then in the field of physics, on an issue concerning hydrostatics (buoyancy force).

Galileo and Ludovico spent three days debating the water and ice issue. Ludovico's basic premise was that ice was the solid form of water, therefore it was denser than water. He argued that buoyancy was "a matter of shape only." "It had nothing to do with density." Ludovico presented a sphere of ebony to the audience. The sphere was placed on the surface of the water, and it began to sink. Then Ludovico took a thin wafer of ebony and placed it on the surface of the water, where it floated. Because the density of both the wafer and the sphere of ebony were the same, Ludovico announced that density had nothing to do with buoyancy and that an object's shape was all that mattered.

Galileo's primary argument for floating ice was correctly based on Archimedes' density theory, wherein an object in water experiences a buoyant force equal to the weight of water it displaces. Because ice is less dense than liquid water, it will always float on liquid water. But Galileo then went too far. Aiming at the main thrust of Ludovico's argument, Galileo said that the shape of an object did not affect whether the object would sink or float. The reason ice floats on water has everything to do with density. Ice is a rare example of a solid that is less dense than its corresponding liquid.

Galileo had ignored the surface tension, however. Surface tension forces can help objects located on a liquid surface resist sinking on the basis of how much of that object is in contact with the liquid's surface. Consider a paper clip: If it is placed flat on the surface of water it can float, but if it is placed on water standing straight up, it sinks. The difference is the higher surface tension force experienced by the paper clip lying flat on the water's surface. So in a way, the shape of an object (in contact with the surface) does contribute to whether it sinks or floats.

The dispute became noisy and inconclusive, and the meeting was brought to a close. The patrons of both Ludovico and Galileo encouraged the two men to write up descriptions of the debate and their arguments, which led to publications of *An Essay on Objects that Float in Water or that Move in It* (Florence, 1612); *A Defence of Galileo's Essay* (1612) and *Considerations concerning Galileo's Essay* (1613 by Ludovico). Both tracts attacked Galileo's theories on the basis of Aristotelian precepts. In 1615, Galileo published a book *Response to the Disagreements of Ser Lodovico delle Colombe and Ser Vincenzo di Grazia against Signor Galileo's Treatise concerning Objects that Lie on Water.*

To commemorate the 400th anniversary of this debate, two dozen researchers met in Florence, Italy, for a week in July 2013 to discuss current unanswered questions in water research at a conference playfully dubbed Aqua Incognita (which can be translated as Water in Disguise or Unknowable Water). This discussion has led to a book edited by Barry Ninham, Pierandrea Lo Nostro [10].

The two water deliberations, some 400 years apart, had similarities: Both were multiday events featuring occasional raucous disagreement about experimental details or theoretical constructs. However, with the hindsight of four centuries, the earlier water debate provides a cautionary tale to water researchers—and in fact all scientists—about the double-edged sword of scientific arrogance.

Four hundred years after the debate, there are still many unresolved questions about water. The fundamental origin in terms of structure and dynamics of its many anomalous properties is still under debate. No model is currently able to reproduce these properties throughout the phase diagram. A four week symposium was held in Nordita, Stockholm, during October 13 and November 07 2014. This program, organized by Lars Pettersson, Anders Nilsson, and Richard Henchman, brought together hundreds experimentalists and theoreticians in strong synergy to explore interpretations and to provide a strong basis for further experimental and theoretical advances towards a unified picture of water. The primary objective of the program is to identify critical aspects of water's anomalous behavior that need to be included in new water models in order to give an overall encompassing agreement with experiments. It also aimed to stimulate further developments of models that can also include perturbations due to ion solvation, hydrophobic interactions as well as describe water at interfaces.

### 1.4 Notes on existing and unknown mechanisms

Currently available mechanisms for density anomalies are mainly focused on the density change in the quasi-solid regime. The mechanism behind the 'regular' process of cooling densification in the liquid I and solid III phase has attracted little attention. The following mechanisms address the freezing expansion in the II phase in terms of supercooled liquid:

1) The mixed-phase scheme [2, 11-17] suggests that a competition between the randomly distributed, 'ice-like' nanoscale fragments, or the ring- or chain-like low-density liquid (LDL), and the tetrahedrally structured high-density liquid (HDL) fragments dictates the volume expansion in the supercooled liquid [13, 18]. Cooling increases the fraction of the LDL phase, and then ice floats. The many-body electronic structure and the non-local van der Waal (vdW) interactions were suggested as possible forces driving volume expansion [19].

The extremely high sensitivity of water to the thermal and experimental conditions evidences that any perturbation can changes the patterns of ice crystal [20] and the cooling rate of water freezing [21]. Masaru Emoto, a Japanese scientist, examined the crystal patterns from water samples subject to simple words like love, thank you, war and hate. He observed the ice crystals under the microscope (see Figure 7 for instance). The samples subjected to love and thank you formed into brilliant crystal shapes. The hate and war samples formed ugly, amorphous shapes.

It is not surprising that harmonious classical music like that of Bhimsen Joshi, Pandit Ravi Shankar, Ali Akbar Khan, Mozart and Beethoven or of harmonious new-age music, and natural sounds of the sea, whales, etc., have a benevolent influence on the patterns of the ice crystal. The opposite has been the case with disharmonious music like heavy metal, sounds like traffic noise and words like "I hate you". Any sound waves or bioelectronics signals of thinking an demotion at different tones or frequencies affect the growth manner of ice crystals.

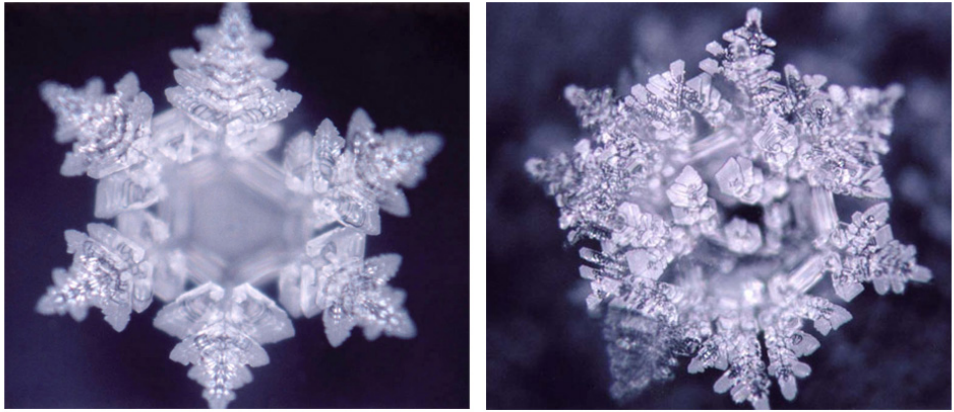

Figure 7 Cooling condition sensitive of the structure patterns of ice crystals exposed to the word "Love" and "Hope" [20].

James Brownridge, a radiation safety officer and nuclear instrumentation specialist at the State University of New York, has conducted over 20 experiments in a 10-year time to obtain true solution to the Mpemba effect – hot water freezes faster [22]. Being subject to the initial temperature, environment, cooling rate, and perturbation, freezing of ice takes different times showing the supercooling effect [21]. Water is so sensitive to the experimental conditions and the external signals that make experiments hardly reproducible. One can imagine what conditions could produce the mixed-phase composed of the LDL and the HDL nanosized fragments in the supercooled phase.

2) The monophase notation [23-27] explains that water contains a homogeneous, three-dimensional, tetrahedrally coordinated structured phase with thermal fluctuation that is not quite random [27, 28]. The monophase model attributes freezing expansion to the O:H-O bond relaxation in length and angle in a fixed yet unclear manner.

In fact, bulk water prefers the homogeneous tetrahedrally- coordinated, fluctuating structure with a supersolid skin of the same geometry but smaller molecular size and larger separation; water at the nanometer scale forms two-phase structure in a core-shell configured bulk and skin [29]. The density oscillates in four-regime over the full temperature range because of the segmental specific heat disparity of the O:H-O bond between oxygen ions. The O:H-O bond angle cooling stretching and its cooperative length relaxation dictate the density change over the full temperature range.

3) The linear correlation model [30] rationalizes that the local density changes homogeneously with the length and angle. Matsumoto used computer simulation to look at ways changing the O:H-O bond network volume: extension of the bonds, change in the containing angle between the bonds, and change in the network topology. He found the O:H-O bond elongation is responsible for thermal expansion, while the angular distortion causes thermal contraction. The network topology does not contribute to volume change. Therefore, the competition between the O:H-O bond angle and its length relaxation determines the density anomalies of water ice.

In fact, O:H-O bond thermal elongation proceeds only in the liquid and in the solid phase but O:H-O bond thermal contraction occurs in the quasi-solid because of the specific heat disparity. The O:H-O bond containing angle is subject to cooling stretching in the solid and the quasi-solid regimes. During cooling in the quasi-solid phase, the H-O bond with the lower specific heat contracts less than the O:H nonbond expands, which lengthens the O:H-O and enlarges the volume, meanwhile, cooling stretches the O:H-O containing angle from 160 up to 167 ° at the least density. Both O:H-O bond elongation and angle stretching contribute to the cooling expansion in the quasi-solid phase, allowing ice to flow. In the solid phase, competition between the θ cooling stretching the O:H-O cooling contraction raises the density slightly compared with the liquid phase where the θ remains constant.

4) The model of two kinds of O:H bond [31, 32] suggests that one kind of stronger and another kind of weaker O:H bond coexist randomly in the ratio of about 2:1. By introducing these two types of O:H bond, Tu and Fang [32] reproduced a number of the anomalies, particularly the thermodynamic properties in the supercooled state. They found that the exchange between the strong and the weak O:H bonds enhance the competition between the open and the collapsed structures of liquid water.

The bulk and skin phase do exist with the stronger O:H bond in the bulk and the weaker in the skin. The volume ration $V_{skin}/V_{bulk}$ between these two kinds of O:H bonds increases with the drop of the droplet size [33].

*1.5    Quantitative evidence*

A number of issues on water density anomalies remain yet unattended [34-37]. Determination of the following attributes is beyond the scope of available models focusing on the phase composition in the supercooling state. Solving the emerging challenges from the perspective of O:H-O bond relaxation forms the subject in this section:

1) Thermal oscillation dynamics of the characteristic phonon frequencies $\omega_x$ and the mass density $\rho$ over the full temperature range
2) The 'regular' process of cooling densification in the liquid and in the solid phases
3) Slightly cooling expansion and $\omega_x$ conservation at extremely low temperatures
4) Correlation between the O 1s thermal entrapment and $\omega_x$ phonon relaxation

1.5.1    Density anomalies and transition temperatures

The $\rho(T)$ profiles for water droplets of 1.4, (Figure 1b) [2] 3.3, 3.9, and 4.4 nm [38] sizes (see Figure 8a) exhibit four regimes of different slopes, transition from I to II, II to III, and III to IV phase occurs at the temperature ranges of 277-315 K(maximal density $\rho_M$ at temperatures close to the melting $T_m$) , 173-258 K(minimal $\rho_m$ at temperature nearby freezing), and 55-80 K ($\Delta\rho$ ~0 drops slightly at cooling)[29]. The transition temperatures vary with the droplet size of water, which is often regarded as size induced "supercooling in freezing" and "superheating at melting".

In the liquid (I, bulk) phase and in the solid (III) phase, $H_2O$ exhibits the normal process of cooling densification at different rates: $d\rho/dT < 0$; $|(d\rho/dT)_I| > |(d\rho/dT)_{III}|$. At the quasi-solid (II) and the solid (IV) phase, volume cooling expansion occurs: $(d\rho/dT)_{II} > 0$; $(d\rho/dT)_{II} \gg (d\rho/dT)_{IV}$ (Figure 8b) [39]. The sixteenth (XVI) phase in the (empty hydrate) cage structure has a density of 0.81 g·cm$^{-3}$. This phase expands slightly when cooling in temperatures below 55 K [40]. This cage structure is mechanically more stable and has at low temperatures larger lattice constants than the filled hydrate, because of the reduction of the effective molecular coordination numbers (CNs) [41]. These observations evidence that neither the H-O nor the O:H undergoes relaxation in length and stiffness but the ∠O:H-O angle is subject to cooling stretching in the IV[th] temperature regime where $\eta_L \approx \eta_H \approx 0$ [5], which results in the slightly cooling expansion.

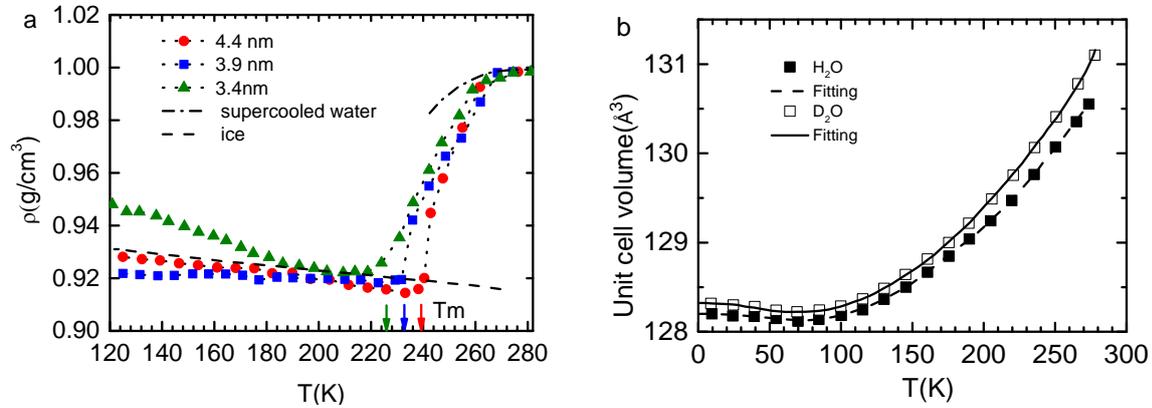

Figure 8. Density $\rho(T)$ profiles of water droplets of different sizes measured using (a) small-angle X-ray scattering [38] and (b) the slight volume expansion happens to $H_2O$ and $D_2O$ at $T \leq 75$ K (IV) [39]. The II-III phase transition takes place at 173-258 K for 1.2 nm droplet [42] and bulk water [3]. Slightly cooling expansion starts from 55-80 K for the cage structured XVI[th] phase [40] and bulk. (Reprinted with permission from [38, 39].)

1.5.2    Bond angle–length relaxation and density oscillation

Figure 2 features the O:H-O bond segmental lengths $d_x$ derived from molecular dynamics (MD) calculations [3] and experimental observations [2]. The MD derivatives show the purely $d_x$ relaxation whose accuracy is subject to numerical algorithm employed. The experimental derivative is a resultant of the $d_x$ relaxation and the $\theta$ relaxation.

Figure 9a shows the MD trajectory snapped at 100, 200 and 300 K temperatures. Figure 9b is the O-O distance as a function of temperature, which agrees quantitatively to measurements in Figure 1b. Agreement between MD and experimental observations asserted that the shortening of the master segments (the part of relatively lower specific heat as denoted with arrows) is always coupled with a lengthening of the slaves during cooling. In the liquid region I and in the solid region III, the O:H nonbond having a lower $\eta_L$ contracts more than the H-O bond elongates, resulting in a net loss of the O-O length. Thus, cooling-driven densification of $H_2O$ takes place in both the liquid and the solid phases. This mechanism differs completely from that conventionally adopted for the standard cooling densification of other 'normal' materials in which only one kind of chemical bond is involved [43].

In contrast, in the II phase, the master and the slave exchange roles. The H-O bond having a lower $\eta_L$ contracts less than the O:H bond expands, producing a net gain in the O-O length and resulting in density loss. Calculations reveal no region IV below 80 K as observed, due to the limitation of the algorithm. Quantitatively, the widening of the angle $\theta$ in Figure 2b contributes consistently to volume expansion. In the liquid phase I, the mean $\theta$ value of 160° remains almost constant, which has little to do with density change but the O:H cooling contraction and H-O elongation dominate.

The snapshots of the MD trajectory in Figure 9a and the MD video in [3] show that the V-shaped H-O-H motifs remain intact at 300 K over the entire duration of recording. This configuration is accompanied by large fluctuations of the $\theta$ and the $d_L$ flashing in this regime but retain the mean of the tetrahedrally-coordinated structure of water molecules [44], even for a single molecule at 5 K temperature [45]. The MD-video in [3] shows that, in the liquid phase, the H and the O attract each other in the O:H interaction, but the O-O repulsion prevents this occurrence. The intact H-O-H motifs move ceaselessly like a complex pendulum because of the high fluctuation and frequent switching the O:H interaction on and off.

In region II, cooling widens $\theta$ from 160° to 167°, which contributes a maximum of +1.75% to the O:H-O bond elongation and about 5% to the volume expansion. The volume expansion due to angle stretching is compatible to O:H-O cooling elongation, resulting in a 9% less density.

In phase III, $\theta$ increases from 167° to 174° and this trend results in a maximal value of -2.76% to the volume contraction in ice. The angle cooling stretching compensates for the O:H-O contraction of bond, which explains why the density gains at a lower rate in the solid phase than it is in the liquid phase. An extrapolation of the θ widening in Figure 2b results in the slight O—O lengthening in region IV where the $d_x$ and its cohesive energy $E_x$ remain constant as $\eta_x \approx 0$, which explains the slight drop in density [39, 40] and the steady $\omega_L$ ($d_L$ and $E_L$) observed at temperatures below 80 K [46, 47]. Therefore, the angle cooling stretching contributes only positively to the density loss in phase II but negatively to density gain in regime III without apparent influence on other physical properties such as the critical temperature for phase transition $T_C$, Oxygen O 1s energy shift $E_{1s}$, and phonon frequencies $\omega_x$ etc.

The O-O distance evolution shown in Figure 9b agrees well in trend with the measured density evolution in the full temperate range, Figure 1b [2]. In ice, the O-O distance is always longer than in water — hence ice floats, without necessary involvement of the mixed-phase configuration. Therefore, the entire process of density oscillation arises from O:H-O bond segmental length relaxation subject to the specific-heat disparity and bond angle cooling stretching.

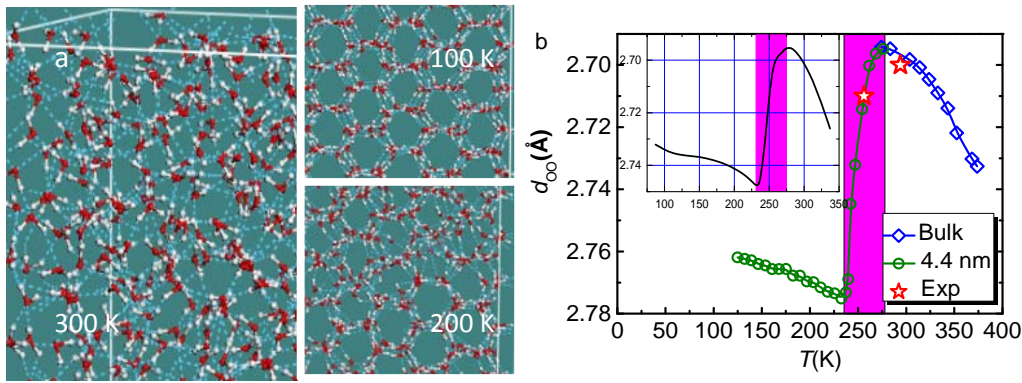

Figure 9. (a) Snapshots of MD trajectory show that the structural order decreases with increasing temperature from 100 to 300 K while the V-shaped $H_2O$ motifs remain intact at 300 K because of the stronger H-O bond (3.97 eV/bond)[29]. (b) O-O distance oscillation profiles derived from measurements [38] and computation (inset) agrees with the measured density trends of water ice except for transition temperatures at 202-258 K [2]. The $d_{OO}(T)$ profile also matches to the measured $d_{OO}$ at 25 and -16.8°C [48] (Reprinted with permission from [3].)

The O-O distance dominates the mass density of water ice in the manner of $\rho \propto (d_{O-O})^{-3} \propto (d_H + d_L)_{//}^{-3}$. The $d_x$ is the projection along O-O without contrbution from the θ contribution, which remains > 160° in all phases [3]. The angle difference between 160° and 180° deviates by only 3% or less to the length scale [3].

When the structures are different, there are other possible volume changes. For example, ice VII has a smaller volume and longer intermolecular distance than ice Ic because the former has double the network of the latter. Ice VII and VIII have similar network connectivities but different crystal symmetries [49]. The transition between these two phases is of the first order [50, 51]. Volume change by such structure variation contributes insignificantly to the O:H-O bond relaxation that dictates the anomalous behavior of water ice.

1.5.3   Phonon stiffness cooperative thermal oscillation

The segmental cohesive energy $E_x$ is inversely proportional to its length $d_x$ and its stiffness (phonon frequency $\omega_x$) follows the relationship:

$$\Delta\omega_x \propto \sqrt{E_x/\mu_x}/d_x \propto \sqrt{(k_x+k_C)/\mu_x}\ .$$

Where the $\mu_x$ is the reduced mass of the H-O and the (H$_2$O):(H$_2$O) vibration dimers. The $k_x$ and $k_C$ are the force constant for the segmental short-range interaction and the inter oxygen repulsion. According to the principle of Fourier transformation, the characteristic phonon spectral peak represents all segments of the same kind disregarding their locations and numbers. The spectrum in full widows gives direct and comprehensive information regarding the cooperative relaxation dynamics of the segmental length, stiffness, and energy of the entire O:H-O bond.

Under any circumstance, the characteristic $\omega_L$ and $\omega_H$ always shift in opposite direction because of the Coulomb repulsion. If one undergoes blue shift, the other does red without any exception. This forms the straightforward yet simple advantage of multifield phonon spectrometrics.  Figure 10 shows the Raman spectra of water droplet of millimetre size cooled from 298 K to 98 K using programmed liquid nitrogen. The spectra show expected three regimes transiting at the quasi-solid phase boundaries of 273 K and 258 K [3]:

1) In the liquid phase I, $T \geq 273$ K, cooling stiffens $\omega_L$ abruptly from 75 to 220 cm$^{-1}$ and softens $\omega_H$ from 3200 to 3140 cm$^{-1}$ with indication of ice forming at 273 K. The cooperative $\omega_x$ shift indicates that cooling shortens and stiffens the O:H bond but lengthens and softens the H-O bond in the liquid phase, which confirms again that the O:H bond cooling contraction dominates O:H-O relaxation in liquid phase.
2) In the phase II, $273 \geq T \geq 258$ K, the situation reverses. Cooling stiffens $\omega_H$ from 3140 to 3150 cm$^{-1}$ and softens $\omega_L$ from 220 to 215 cm$^{-1}$ (see the shaded areas). Consistent with the Raman $\omega_H$ shift measured at temperatures around 273 K [52, 53], the cooperative shift of $\omega_x$ confirms the switching of the master and the slave roles of the O:H and H-O during freezing; H-O contraction dominates in this quasi-solid phase.
3) In the solid phase III, T ≤ 258 K, the master-slave role reverts to its behavior in the liquid region,

albeit with a different relaxation rate. Cooling from 258 to 98 K stiffens $\omega_L$ from 215 to 230 cm$^{-1}$ and softens $\omega_H$ from 3150 to 3100 cm$^{-1}$ as it cools. Earlier Raman spectroscopy revealed that the $\omega_L$ for bulk ice and $D_2O$ drops monotonically with the rise of temperature and the data fluctuates at 260±10 K [46]. The supplementary peaks at about 300 and 3450 cm$^{-1}$ change insignificantly with temperature; the skin $\omega_H$ of about 3450 cm$^{-1}$ in water and ice is indeed thermally insensitive [54]. The cooling softening of $\omega_H$ agrees with that measured using IR spectroscopy of ice clusters of 8–150 nm size [55]. When the temperature drops from 209 to 30 K, $\omega_H$ shifts from 3253 to 3218 cm$^{-1}$.

4) Figure 11 shows that both the $\omega_H$ and the $\omega_L$ remain almost constant at $T < 60$ K [55]. Using IR spectroscopy, Medcraft et al [47] measured the size- and temperature-dependence of $\omega_L$ in the temperature range 4–190 K. They found that heating softens the $\omega_L$ at $T > 80$ K but the $\omega_L$ remains almost unchanged below 60 K. This observation evidences that neither the length nor the stiffness or energy of these two segments change in this temperature regime IV because of their extremely low specific heat ($\eta_x \approx 0$).

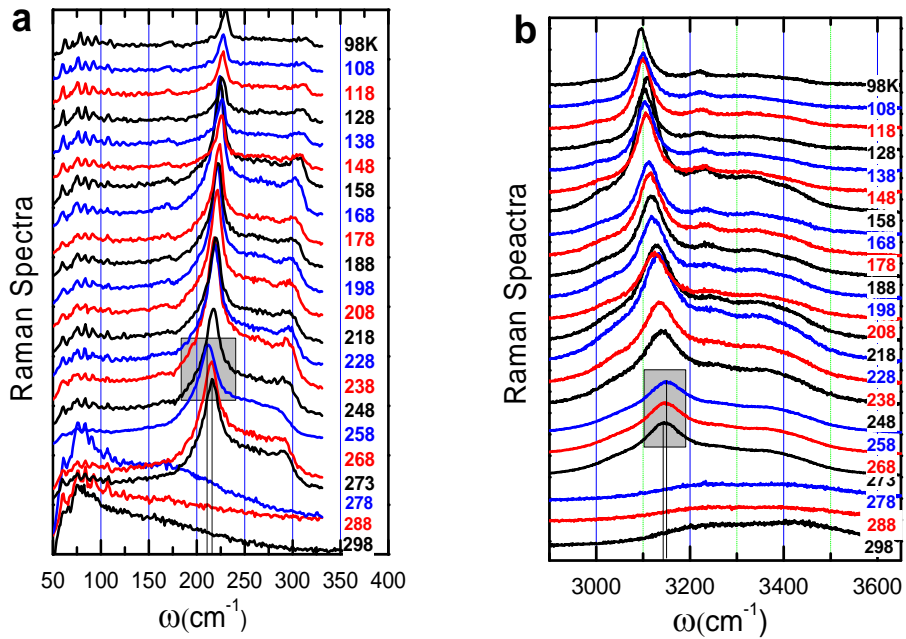

Figure 10. Temperature-dependent Raman shifts of (a) $\omega_L < 300$ cm$^{-1}$ and (b) $\omega_H > 3000$ cm$^{-1}$ in the temperature regions of $T > 273$ K, $273 \geq T \geq 258$ K, and $T < 258$ K, agreeing the O:H-O length, density, and stiffness cooperative oscillation over the full temperature range (reprinted with permission from [3]).

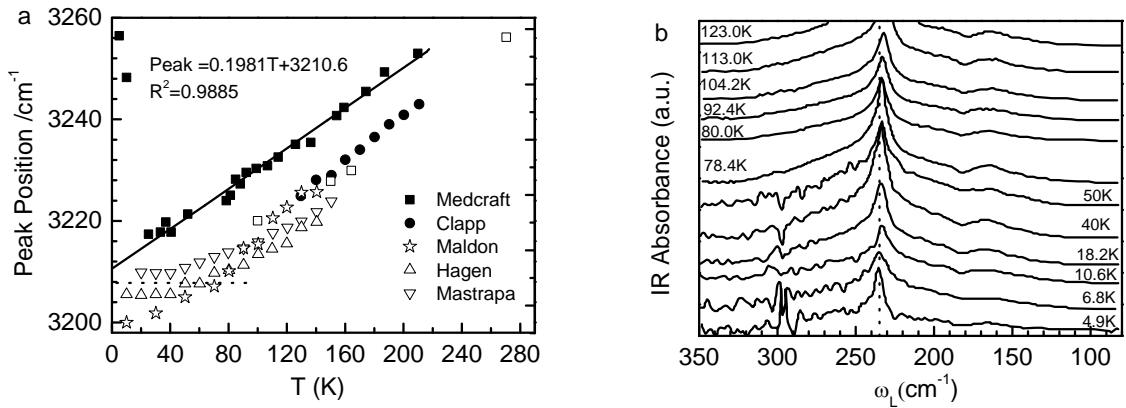

Figure 11. Insignificant shift of (a) the $\omega_H$ and (b) the $\omega_L$ at $T \leq 60$ K This indicates that $\eta_x \cong 0$ almost silences the O:H-O bond length and stiffness in this temperature regime [3]. Broken lines guide viewing. (reprinted with permission from [47, 55] and references therein).

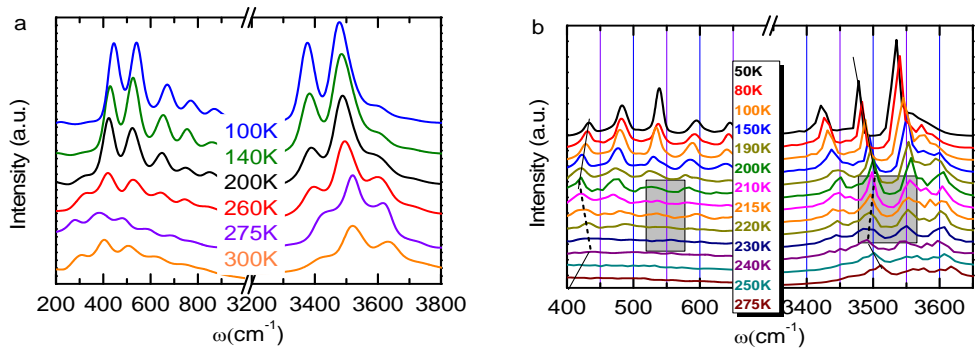

Figure 12. Temperature-dependent power spectra of $H_2O$. (a) Splitting of the high-frequency peaks at 260 K indicates the transition from water to ice at 200–260 K. (b) Phonon oscillation (indicated with hatched lines) holds the same trend as that of Raman measurements in

Figure 12 shows the $T$-dependent power spectra of $H_2O$ derived from MD calculations. The splitting of the high-frequency peaks at 260 K indicates the transition from water to ice at 200–260 K. The three-region phonon thermal oscillation is the same as the measurements in. Figure 13 compares the measured and the calculated phonon thermal relaxation dynamics. As expected, $\omega_L$ stiffening (softening) always couples with $\omega_H$ softening (stiffening) in all three regions, which evidence the cooperative relaxation of the O:H-O bond in these regions.

Offsets of the calculated $\varpi_L$ by -200 cm$^{-1}$ and $\varpi_H$ by -400 cm$^{-1}$ compared to experiments suggest the presence of artifacts in the MD algorithm that deals inadequately with the ultra-short-range interactions.

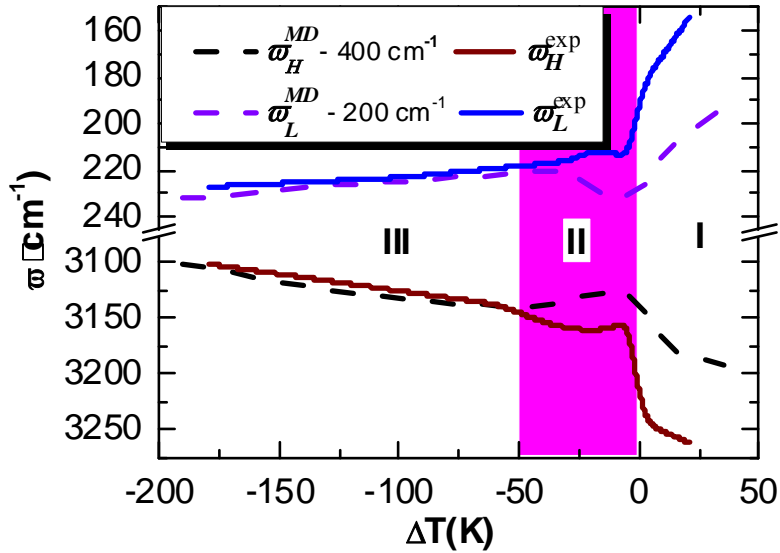

Figure 13. Comparison of the measured (solid lines) and the calculated (broken lines) phonon relaxation dynamics. Indicated 200/400 cm$^{-1}$ offsets of the calculated $\varpi_x$ match observations. (Reprinted with permission from [3].)

1.5.4   H-O phonon relaxation

Phonons of 'normal' materials undergo heat softening because of the thermal lengthening and softening of all bonds involved [56-63]. Figure 14 shows, however, heating stiffens the stiffer $\varpi_H$ phonons of water [64-70] and ice [35, 36, 52, 67, 71]. The $\varpi_H$ increases abruptly to saturation at evaporation and then remains constant in the vapor phase composed of monomers unless the water undergoes superheating (Figure 14a). The spectral shape changes from 3150 cm$^{-1}$ dominance to 3250 cm$^{-1}$ dominance when the temperature changes from -10 to 10 °C with strengthening of the 3450 cm$^{-1}$ skin features, as shown in Figure 14b.

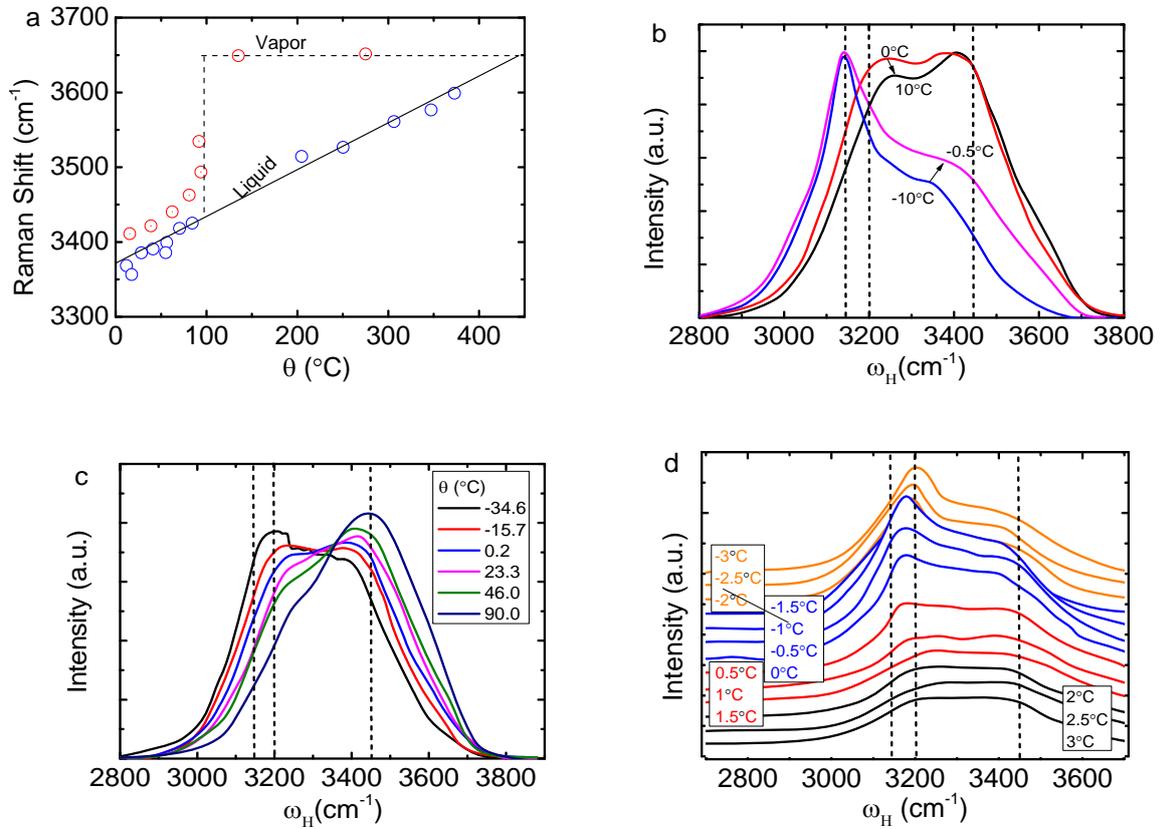

Figure 14. Thermal $\omega_H$ stiffening for bulk water in the temperature ranges from (a) 0 to 300°C, and (b) -10 to 10°C. (c) Thermal $\omega_H$ stiffening for supercooled water droplets in the range from -34.6 to 90°C. (d) The $\omega_H$ transits at 0°C from thermal-stiffening to cooling-stiffening. The $\omega_H$ in (a) is unchanged in the vapor phase (red circles) at 3650 cm$^{-1}$; blue circles show linear dependence of $\omega_H$ on temperature in the superheating liquid. Hatched vertical lines indicate $\omega_H$ at 3450 cm$^{-1}$ for the skin, 3200 cm$^{-1}$ for bulk water, and 3150 cm$^{-1}$ for bulk ice. (Reprinted with permission from [53, 64, 67, 72].)

Figure 14c shows the $\omega_H$ evolution from 3200 cm$^{-1}$ dominance to 3400 cm$^{-1}$ dominance as the free-standing water droplets changes from the supercooling state at -34.6°C to 90.0°C. Thermal $\omega_H$ stiffening proceeds consistently throughout the liquid phase [72]. Marechal [73] observed that thermal $\omega_H$ stiffening and $\omega_L$ softening happen simultaneously, not only in liquid H$_2$O but also to liquid D$_2$O, despite an offset in the characteristic peak. However, the $\omega_H$ transition at 0°C from thermal stiffening to thermal softening, as shown in Figure 14d [67], and the $\omega_x$ coupling in Figure 10b have hitherto received little attention.

1.5.5  Oxygen $\Delta E_{1s}$ versus $\Delta\omega_H$ thermal relaxation

The energy shift of the O 1s energy level from that of an isolated O atom measured using x-ray photoelectron spectroscopy (XPS) is proportional to the H-O bond energy (4.0 eV) as the O:H nonbond energy (0.10 eV) contribution is negligibly small. Information given by the near edge absorption fine structure (NEXAFS) and the X-ray emission spectroscopy (XES), as shown in Figure 15, is much more complicated as both the O 1s (bottom) and the upper occupied and unoccupied levels are subject to shift in different amounts [74]. Electron spectrometrics provides information on the H-O bond relaxation in energy and the associated bonding charge entrapment.

It is necessary to point out that the often used electron spectrometrics and diffraction methods are subject to limitation in water research. Electron spectrometrics provides little information about the O:H nonbond relaxation in energy because its perturbation to the Hamiltonian is only 0.1/4.0 < 3% compared to the contribution of the H-O bond. The O-O pairing distribution function obtained from XRD or neutron diffraction gives no details about the O:H and H-O cooperative length change. The phonon spectrometrics over the full frequency windows presents most comprehensive information of the O:H-O bond cooperative relaxation dynamics.

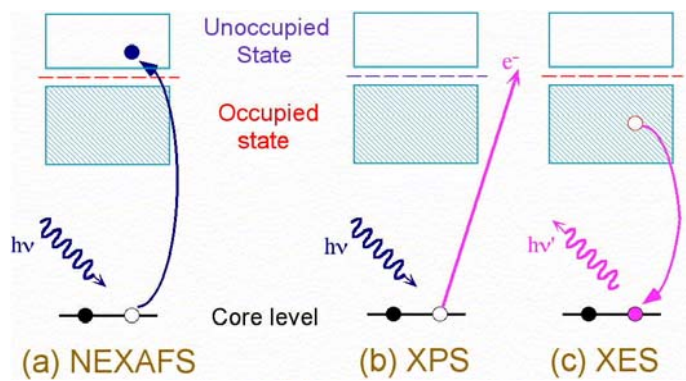

Figure 15 Principles for the electronic spectroscopy techniques. O 1s electron absorbs energy in the NEXAFS process and transits to the upper unoccupied levels. After thermalization the excited electron transits to the O 1s level and emits the XES energy. The XPS exited provides the isolated O 1s level binding energy while the rest involves two levels [74].

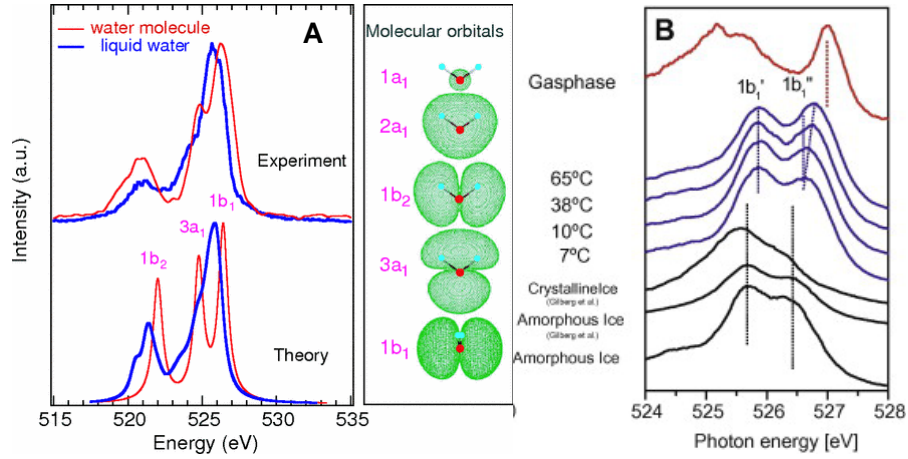

Figure 16. (A) The O 1s orbital (side panel) energies of molecules and liquid water and (B) the O 1s XES spectra of vapor, liquid water, and amorphous and crystalline ice at different temperatures, with an energy scale displaying the 1b$_1$ orbitals. The 1b$_1$ peak splits into a doublet of 1b$_1'$(~525.5 eV) and 1b$_1''$ (~526.5 eV) corresponding the $\omega_H$ for the bulk (3200 cm$^{-1}$) and skin (3400 cm$^{-1}$) respectively, which undergo thermal entrapment/stiffening consistently for crystals but thermal softening for amorphous ice transiting to crystals. (Reprinted with permission from [75, 76].)

Figure 16 shows that heating deepens the O 1s energy in different phases towards that of gaseous molecules unless transition from amorphous to crystal [75, 76]. Mechanisms for the O 1s thermal entrapment are debated as consequence of the mixed-phase configuration, that is, ordered tetrahedral and distorted O:H-O bonded networks, with provision of the mixed-structure phase [18, 77].

In fact, the following correlates the O 1s energy shift and the H-O phonon frequency shift $\Delta\omega_H$ [29]:

$$\left(d_H \Delta\omega_H\right)^2 \cong \Delta E_{1s}$$

This relation indicates that both the $\Delta E_{1s}$ and the $\Delta\omega_H$ always shift in the same direction, at different rates, when the specimen is excited. Therefore, the 1b$_1''$ peak corresponds to the skin $\omega_H$ at 3450 cm$^{-1}$, and the 1b$_1'$ to the bulk $\omega_H$ at 3200 cm$^{-1}$ for water (see Figure 16). The O 1s goes deeper in the crystal, and the $\omega_H$ shifts consistently higher at heating, because heating shortens and stiffens the H-O bond. When ice transits from amorphous to crystal, the trend is opposite, (see Figure 17) [78]. The $\Delta E_{1s}$ will undergo thermal oscillation but its measurement in ultra-high vacuum is very difficult.

*1.6  Indication on annealing of amorphous ice*

Thermal annealing of low-density amorphous ice from 80 to 155 K softens $\omega_H$ from 3120 to 3080 cm$^{-1}$[78], which is counter to the trend of $\omega_H$ heating stiffening in ice-VIII crystals. Thermal relaxation increases the structural order of the amorphous state on more extended length scales as the average O-O distance becomes shorter with narrower distribution.

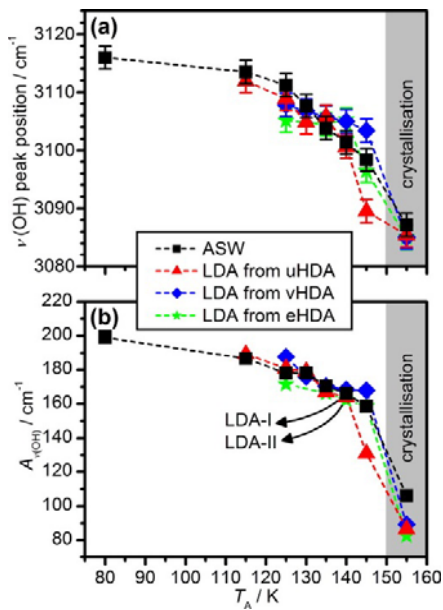

Figure 17 Annealing temperature dependence of (a) the $\omega_H$ peak position and (b) the integrated peak areas. L/HDA represents the low/high-density amorphous ice and ASW for amorphous solid water (Reprinted with permission from [78].)

Clearly, heating softens the $\omega_H$ of amorphous ice rather than stiffening it, as it occurs in crystalline ice. The $\omega_H$ redshift indicates H-O bond elongation. Molecular undercoordination shortens the H-O bond nearby defects [41], which distributes randomly in the amorphous phase. Annealing removes the defect and relaxes the $\omega_H$ towards crystallization with H-O elongation. Therefore, $\omega_H$ redshift occurs in amorphous ice upon annealing, which is within expectation of the bond order-length-strength (BOLS) notation [29].

*1.7  Indication on Isotope effect*

Figure 18 shows that the isotope (Deuterium) has two effects on the IR spectrum of ordinary H₂O [73]. One is the intensity attenuation of all peaks and the other is the general phonon softening [73]. However, $\omega_x$ maintains the shift trend due to heating - $\omega_H$ stiffening and $\omega_L$ softening. Knowledge developed herewith clarifies the mechanism of this isotope effect.

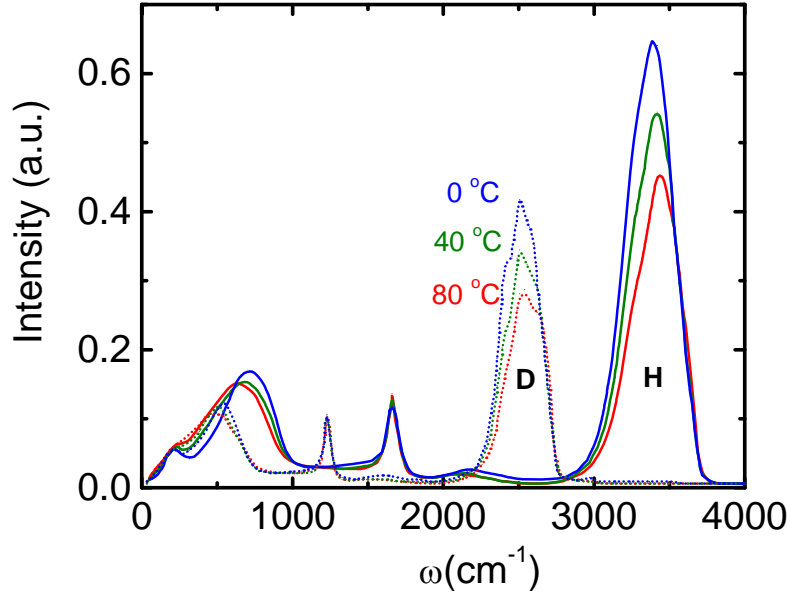

Figure 18. IR spectra of 1 μm-thick ordinary water (H) and heavy water (D) reveal that isotope D attenuates the intensity and softens all phonons (redshift) of the H₂O in general, although the trend of thermal $\omega_H$ stiffening and $\omega_L$ thermal softening remain. (Reprinted with permission from [73].)

The effective mass reduction due to isotope dictates the phonon relaxation. The isotope contributes only to reducing the $\mu(m_1, m_2) = m_1 m_2 / (m_1 + m_2)$ in the $\omega_x \propto (E_x/\mu_x)^{1/2}/d_x$ expression. Considering the mass difference, both vibration modes shift their relative frequencies compared to ordinary water in the following manner:

$$\frac{\Delta\omega_{xH}}{\Delta\omega_{xD}} \cong \left(\frac{\mu_D(m_1,m_2)}{\mu_H}\right)^{1/2} = \begin{cases} \left[\dfrac{\mu_D(2,16)}{\mu_H(1,16)}\right]^{1/2} = (17/9)^{1/2} = 1.374 \quad (\omega_H) \\ \left[\dfrac{\mu_D(20,20)}{\mu_H(16,16)}\right]^{1/2} = (5/4)^{1/2} = 1.180 \quad (\omega_L) \end{cases},$$

(1)

where, for intramolecular H-O vibration, $m_1$ is the mass of H (1 atomic unit) or D (2 atomic units); and $m_2$ is the mass of O (16 units); and for intermolecular (H$_2$O):(H$_2$O) vibration, $m_1 = m_2$ is the mass of 2H + O (18 units) or 2D + O (20 units). Measurements shown in Figure 18 yield the following:

$$\frac{\Delta\omega_{xH}}{\Delta\omega_{xD}} \approx \begin{cases} 3400/2500 = 1.36 & (\omega_H) \\ 1620/1200 = 1.35 & (\omega_B) \\ 750/500 = 1.50 & (\omega_L) \end{cases}$$

(2)

The difference between the numerical derivatives in Eq. (1) and measurements in Eq. (2) arises mainly from Coulomb coupling, particularly for $\omega_L$. Such a first-order approximation is effective for describing the isotopic effect on the phonon relaxation dynamics of $\omega_x$. Therefore, the addition of the isotope softens all the phonons by mediating the effective mass of the coupled oscillators in addition to the quantum effect that may play a certain role. The peak intensities in the isotope are also lower because low-frequency vibrations enhance phonon scattering.

*1.8    Indication on general thermal contraction*

The vast majority of materials have a positive coefficient of thermal expansion ($\alpha(\theta) > 0$) and their volume increases on heating. There is also another very large number of materials that display the opposite behavior: their volume contracts on heating, that is, they have a negative thermal expansion (NTE) coefficient [79-82]. A typical specimen is cubic ZrW$_2$O$_8$ that contracts over a temperature range exceeding 1000 K [83]. NTE also appears in diamond, silicon and germanium at very low temperatures (< 100 K) [84], and in glass in the titania-silicate family, Kevlar fiber, carbon fibers, anisotropic Invar Fe–Ni alloys, and certain kinds of molecular networks at room temperature. The NTE of graphite [85], graphene oxide paper [86], and ZrWO$_3$ [83] all share the NTE attribute of water at freezing, see Figure 19. NTE materials may be combined with other materials with a positive thermal expansion coefficient to fabricate composites having an overall zero thermal expansion (ZTE). ZTE materials are useful because they do not undergo thermal shock on rapid heating or cooling.

The typical model that explain the NTE effect suggests that NTE arises from the transverse thermal vibrations of the bridging oxygen in the M-O-M linkages inside ZrW$_2$O$_8$, HfW$_2$O$_8$, SC$_2$W$_3$O$_{12}$, AlPO$_{4-17}$, and faujasite-SiO$_2$ [87, 88]. The phonon modes (centered around 30 meV or 3200 cm$^{-1}$) [89] can propagate without distorting the WO$_4$ tetrahedron or the ZrO$_6$ octahedron, termed the 'rigid-unit mode'. The rigid-unit mode also accounts for the structural phase transition of ZrW$_2$O$_8$ and ZrV$_2$O$_7$ [90].

Extending the NTE mechanism for the NTE in the quasi-solid phase provides an atomistic view on the NTE in general. The NTE results from the involvement of at least two kinds of coupled, short-range interactions and the associated specific-heat disparity. In the instance of graphite, the (0001) intralayer covalent bond and the interlayer vdW interactions may play certain roles, in much the same way as the O:H-O bond does in water. O, N and F all create lone pairs of electrons upon reaction, which create the weaker short-range nonbonding interaction. Phonon spectroscopes have the capability to monitor the relaxation process easily and directly, as they do for water.

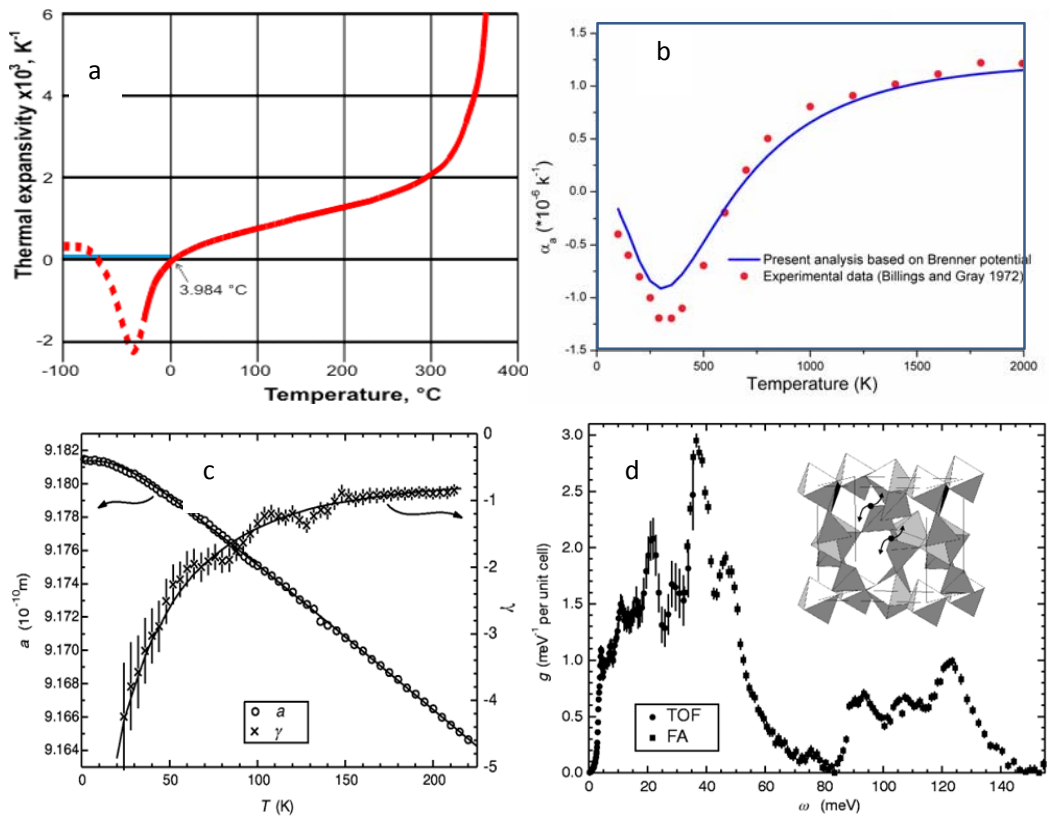

Figure 19. The NTE of (a) $H_2O$; (b) graphite; and (c) $ZrW_2O_8$ with thermal expansion coefficient $\alpha$ (open circles) and Grüneisen parameter $\gamma = 3\alpha B/C_v$ (crosses), where $B$ is the bulk modulus and $C_v$ is the specific heat at constant volume; (d) shows the associated phonon spectrum measured at $T = 300$ K. The inset illustrates the 'rigid rotation model' model (reprinted with permission from [85, 89, 91]). These NTEs share the same behavior as water freezing, but at different temperature ranges, which is evidence of the essentiality of two types of coupled short-range interactions with specific-heat disparity to these materials.

*1.9    Summary*

Consistency in the η, ρ, $d_x$, $\omega_x$ four-region oscillation evidence that the coupled O:H-O bond oscillator pair describes adequately the true situation of water and ice when cooling or heating. Consistency of numerical and experimental observations verifies the following:

1) Inter-oxygen repulsion and the segmental specific-heat disparity of the O:H-O bond govern the change in the angle, length, and stiffness of the segmented O:H-O bond, and the oscillation of the mass density and the phonon-frequency of water ice over the full temperature range.
2) The segment with relatively lower specific heat contracts and drives the O:H-O bond cooling relaxation. Cooling stretching of the ∠O:H-O angle contributes positively to volume expansion in the quasi-solid phase but it contributes negatively to cooling densification in the solid phase. Angle relaxation has no direct influence on the physical properties, with the exception of mass density.
3) In the liquid and solid phases, the O:H bond contracts more than the H-O bond elongates, resulting in the cooling densification of water and ice. This mechanism is completely different from the process experienced by other 'normal' materials when only one type of chemical bond is involved.
4) In the quasi-solid phase, H-O bond contracts less than the O:H bond lengthens, resulting in volume expansion during freezing. Stretching of the O:H-O bond angle lowers the density slightly at $T$ < 80 K as the length and energy of the O:H-O conserve.
5) The O-O distance is larger in ice than it is in water, and therefore ice floats.
6) Understanding clarifies the NTE of other materials and the transition from amorphous to crystal ice under thermal annealing.

1.  H. Perlman. *Ice is less dense than water*. 2014.
2.  F. Mallamace, C. Branca, M. Broccio, C. Corsaro, C.Y. Mou, and S.H. Chen, *The anomalous behavior of the density of water in the range 30 K < T < 373 K.* PNAS, 2007. **104**(47): 18387-18391.
3.  C.Q. Sun, X. Zhang, X. Fu, W. Zheng, J.-l. Kuo, Y. Zhou, Z. Shen, and J. Zhou, *Density and phonon-stiffness anomalies of water and ice in the full temperature range.* J Phys Chem Lett, 2013. **4**: 3238-3244.
4.  Y. Huang, X. Zhang, Z. Ma, Y. Zhou, J. Zhou, W. Zheng, and C.Q. Sun, *Size, separation, structure order, and mass density of molecules packing in water and ice.* Scientific Reports, 2013. **3**: 3005.
5.  C. Lister, *On the penetration of water into hot rock.* Geophysical Journal International, 1974. **39**(3): 465-509.
6.  F. Van der Paauw, *Effect of winter rainfall on the amount of nitrogen available to crops.* Plant and Soil, 1962. **16**(3): 361-380.
7.  A. Pattero. *Sea level rise - National geographic*. 2008; Available from: http://ocean.nationalgeographic.com/ocean/critical-issues-sea-level-rise/.
8.  R. Winkelmann, A. Levermann, M.A. Martin, and K. Frieler, *Increased future ice discharge from Antarctica owing to higher snowfall.* Nature, 2012. **492**(7428): 239-242.
9.  S. Everts, *Galileo On Ice.* Chemical Engineering News, 2013. **91**(34): 28-29.
10. P.L. Nostro and B.W. Ninham, *Aqua Incognita: Why Ice Floats on Water and Galileo 400 Years on* 2014: Connor Court Publishing Pty Ltd.

11. A.J. Stone, *Water from First Principles.* Science, 2007. **315**(5816): 1228-1229.
12. K. Stokely, M.G. Mazza, H.E. Stanley, and G. Franzese, *Effect of hydrogen bond cooperativity on the behavior of water.* PNAS, 2010. **107**(4): 1301-1306.
13. C. Huang, K.T. Wikfeldt, T. Tokushima, D. Nordlund, Y. Harada, U. Bergmann, M. Niebuhr, T.M. Weiss, Y. Horikawa, M. Leetmaa, M.P. Ljungberg, O. Takahashi, A. Lenz, L. Ojamäe, A.P. Lyubartsev, S. Shin, L.G.M. Pettersson, and A. Nilsson, *The inhomogeneous structure of water at ambient conditions.* Proceedings of the National Academy of Sciences, 2009. **106**(36): 15214-15218.
14. F. Mallamace, M. Broccio, C. Corsaro, A. Faraone, D. Majolino, V. Venuti, L. Liu, C.Y. Mou, and S.H. Chen, *Evidence of the existence of the low-density liquid phase in supercooled, confined water.* PNAS, 2007. **104**(2): 424-428.
15. O. Mishima and H.E. Stanley, *The relationship between liquid, supercooled and glassy water.* Nature, 1998. **396**(6709): 329-335.
16. E.B. Moore and V. Molinero, *Structural transformation in supercooled water controls the crystallization rate of ice.* Nature, 2011. **479**(7374): 506-508.
17. V. Molinero and E.B. Moore, *Water Modeled As an Intermediate Element between Carbon and Silicon.* J. Phys. Chem. B, 2009. **113**(13): 4008-4016.
18. P. Wernet, D. Nordlund, U. Bergmann, M. Cavalleri, M. Odelius, H. Ogasawara, L.A. Naslund, T.K. Hirsch, L. Ojamae, P. Glatzel, L.G.M. Pettersson, and A. Nilsson, *The structure of the first coordination shell in liquid water.* Science, 2004. **304**(5673): 995-999.
19. A. Nilsson, C. Huang, and L.G.M. Pettersson, *Fluctuations in ambient water.* J. Mol. Liq., 2012. **176**: 2-16.
20. M. Emoto and E. Puttick, *The healing power of water* 2007: Hay House, Incorporated.
21. J.D. Brownridge, *When does hot water freeze faster then cold water? A search for the Mpemba effect.* Am J Phys, 2011. **79**(1): 78.
22. E.B. Mpemba and D.G. Osborne, *Cool?* Phys. Educ., 1979. **14**: 410-413.
23. G.N.I. Clark, C.D. Cappa, J.D. Smith, R.J. Saykally, and T. Head-Gordon, *The structure of ambient water.* Mol. Phys., 2010. **108**(11): 1415-1433.
24. A.K. Soper, J. Teixeira, and T. Head-Gordon, *Is ambient water inhomogeneous on the nanometer-length scale?* PNAS, 2010. **107**(12): E44-E44.
25. T. Head-Gordon and M.E. Johnson, *Tetrahedral structure or chains for liquid water.* PNAS, 2006. **103**(21): 7973-7977.
26. G.N. Clark, G.L. Hura, J. Teixeira, A.K. Soper, and T. Head-Gordon, *Small-angle scattering and the structure of ambient liquid water.* PNAS, 2010. **107**(32): 14003-14007.
27. V. Petkov, Y. Ren, and M. Suchomel, *Molecular arrangement in water: random but not quite.* J Phys Condens Matter, 2012. **24**(15): 155102.
28. N.J. English and J.S. Tse, *Density Fluctuations in Liquid Water.* Phys. Rev. Lett., 2011. **106**(3): 037801.
29. Y. Huang, X. Zhang, Z. Ma, Y. Zhou, W. Zheng, J. Zhou, and C.Q. Sun, *Hydrogen-bond relaxation dynamics: resolving mysteries of water ice.* Coord. Chem. Rev., 2015. **285**: 109-165.
30. M. Matsumoto, *Why Does Water Expand When It Cools?* Phys. Rev. Lett., 2009. **103**(1): 017801.
31. J.C. Li and A.I. Kolesnikov, *Neutron spectroscopic investigation of dynamics of water ice.* J. Mol. Liq., 2002. **100**(1): 1-39.
32. Y.S. Tu and H.P. Fang, *Anomalies of liquid water at low temperature due to two types of hydrogen bonds.* Phys. Rev. E, 2009. **79**(1): 016707.
33. X. Zhang, Y. Huang, Z. Ma, Y. Zhou, W. Zheng, J. Zhou, and C.Q. Sun, *A common supersolid skin covering both water and ice.* PCCP, 2014. **16**(42): 22987-22994.
34. Y. Yoshimura, S.T. Stewart, M. Somayazulu, H. Mao, and R.J. Hemley, *High-pressure x-ray diffraction and Raman spectroscopy of ice VIII.* J Chem Phys, 2006. **124**(2): 024502.
35. Y. Yoshimura, S.T. Stewart, M. Somayazulu, H.K. Mao, and R.J. Hemley, *Convergent Raman Features in High Density Amorphous Ice, Ice VII, and Ice VIII under Pressure.* J. Phys. Chem. B, 2011. **115**(14): 3756-3760.
36. Y. Yoshimura, S.T. Stewart, H.K. Mao, and R.J. Hemley, *In situ Raman spectroscopy of low-temperature/high-pressure transformations of H2O.* J Chem Phys, 2007. **126**(17): 174505.
37. M. Song, H. Yamawaki, H. Fujihisa, M. Sakashita, and K. Aoki, *Infrared investigation on ice VIII and the phase diagram of dense ices.* Phys. Rev. B, 2003. **68**(1): 014106.

38. M. Erko, D. Wallacher, A. Hoell, T. Hauss, I. Zizak, and O. Paris, *Density minimum of confined water at low temperatures: a combined study by small-angle scattering of X-rays and neutrons.* PCCP, 2012. **14**(11): 3852-3858.
39. K. Rottger, A. Endriss, J. Ihringer, S. Doyle, and W.F. Kuhs, *Lattice-constants and thermal-expansion of H2O and D2O Ice ih between 10 and 265 K.* Acta Crystallographica B, 1994. **50**: 644-648.
40. A. Falenty, T.C. Hansen, and W.F. Kuhs, *Formation and properties of ice XVI obtained by emptying a type sII clathrate hydrate.* Nature, 2014. **516**(7530): 231-233.
41. C.Q. Sun, X. Zhang, J. Zhou, Y. Huang, Y. Zhou, and W. Zheng, *Density, Elasticity, and Stability Anomalies of Water Molecules with Fewer than Four Neighbors.* J Phys Chem Lett, 2013. **4**: 2565-2570.
42. F.G. Alabarse, J. Haines, O. Cambon, C. Levelut, D. Bourgogne, A. Haidoux, D. Granier, and B. Coasne, *Freezing of Water Confined at the Nanoscale.* Phys. Rev. Lett., 2012. **109**(3): 035701.
43. M.X. Gu, Y.C. Zhou, and C.Q. Sun, *Local bond average for the thermally induced lattice expansion.* J. Phys. Chem. B, 2008. **112**(27): 7992-7995.
44. T.D. Kuhne and R.Z. Khaliullin, *Electronic signature of the instantaneous asymmetry in the first coordination shell of liquid water.* Nat Commun, 2013. **4**: 1450.
45. J. Guo, X. Meng, J. Chen, J. Peng, J. Sheng, X.Z. Li, L. Xu, J.R. Shi, E. Wang, and Y. Jiang, *Real-space imaging of interfacial water with submolecular resolution-Supp.* Nat. Mater., 2014.
46. G.P. Johari, H.A.M. Chew, and T.C. Sivakumar, *Effect of temperature and pressure on translational lattice vibrations and permittivity of ice.* J Chem Phys, 1984. **80**(10): 5163.
47. C. Medcraft, D. McNaughton, C.D. Thompson, D. Appadoo, S. Bauerecker, and E.G. Robertson, *Size and Temperature Dependence in the Far-Ir Spectra of Water Ice Particles.* The Astrophysical Journal, 2012. **758**(1): 17.
48. U. Bergmann, A. Di Cicco, P. Wernet, E. Principi, P. Glatzel, and A. Nilsson, *Nearest-neighbor oxygen distances in liquid water and ice observed by x-ray Raman based extended x-ray absorption fine structure.* J Chem Phys, 2007. **127**(17): 174504.
49. V.F. Petrenko and R.W. Whitworth, *Physics of ice*1999: Clarendon Press.
50. P. Pruzan, J.C. Chervin, and B. Canny, *Determination of the d2o ice vii-viii transition line by raman-scattering up to 51 gpa.* J. Chem. Phys., 1992. **97**(1): 718-721.
51. P. Pruzan, J.C. Chervin, and B. Canny, *Stability domain of the ice-VIII proton-ordered phase at very high-pressure and low-temperature.* J Chem Phys, 1993. **99**(12): 9842-9846.
52. I. Durickovic, R. Claverie, P. Bourson, M. Marchetti, J.M. Chassot, and M.D. Fontana, *Water-ice phase transition probed by Raman spectroscopy.* Journal of Raman Spectroscopy, 2011. **42**(6): 1408-1412.
53. X. Xue, Z.-Z. He, and J. Liu, *Detection of water-ice phase transition based on Raman spectrum.* Journal of Raman Spectroscopy, 2013. **44**(7): 1045-1048.
54. T.F. Kahan, J.P. Reid, and D.J. Donaldson, *Spectroscopic probes of the quasi-liquid layer on ice.* J Phys Chem A, 2007. **111**(43): 11006-11012.
55. C. Medcraft, D. McNaughton, C.D. Thompson, D.R.T. Appadoo, S. Bauerecker, and E.G. Robertson, *Water ice nanoparticles: size and temperature effects on the mid-infrared spectrum.* PCCP, 2013. **15**(10): 3630-3639.
56. I. Calizo, A.A. Balandin, W. Bao, F. Miao, and C.N. Lau, *Temperature Dependence of the Raman Spectra of Graphene and Graphene Multilayers.* Nano Lett., 2007. **7**(9): 2645-2649.
57. X.X. Yang, J.W. Li, Z.F. Zhou, Y. Wang, L.W. Yang, W.T. Zheng, and C.Q. Sun, *Raman spectroscopic determination of the length, strength, compressibility, Debye temperature, elasticity, and force constant of the C-C bond in graphene.* Nanoscale, 2012. **4**(2): 502-510.
58. H.Q. Zhou, C.Y. Qiu, H.C. Yang, F. Yu, M.J. Chen, L.J. Hu, Y.J. Guo, and L.F. Sun, *Raman spectra and temperature-dependent Raman scattering of carbon nanoscrolls.* Chem. Phys. Lett., 2011. **501**(4-6): 475-479.
59. X.X. Yang, J.W. Li, Z.F. Zhou, Y. Wang, W.T. Zheng, and C.Q. Sun, *Frequency response of graphene phonons to heating and compression.* Appl. Phys. Lett., 2011. **99**(13): 133108.
60. J.W. Li, L.W. Yang, Z.F. Zhou, X.J. Liu, G.F. Xie, Y. Pan, and C.Q. Sun, *Mechanically Stiffened and Thermally Softened Raman Modes of ZnO Crystal.* J. Phys. Chem. B, 2010. **114**(4): 1648-1651.
61. M.X. Gu, Y.C. Zhou, L.K. Pan, Z. Sun, S.Z. Wang, and C.Q. Sun, *Temperature dependence of the elastic and vibronic behavior of Si, Ge, and diamond crystals.* J. Appl. Phys., 2007. **102**(8): 083524.
62. M.X. Gu, L.K. Pan, T.C.A. Yeung, B.K. Tay, and C.Q. Sun, *Atomistic origin of the thermally driven softening of Raman optical phonons in group III nitrides.* J. Phys. Chem. C, 2007. **111**(36): 13606-13610.

63. M.X. Gu, L.K. Pan, B.K. Tay, and C.Q. Sun, *Atomistic origin and temperature dependence of Raman optical redshift in nanostructures: a broken bond rule.* Journal of Raman Spectroscopy, 2007. **38**(6): 780-788.
64. P.C. Cross, J. Burnham, and P.A. Leighton, *The Raman spectrum and the structure of water.* J Am Chem Soc, 1937. **59**: 1134-1147.
65. J.D. Smith, C.D. Cappa, K.R. Wilson, R.C. Cohen, P.L. Geissler, and R.J. Saykally, *Unified description of temperature-dependent hydrogen-bond rearrangements in liquid water.* PNAS, 2005. **102**(40): 14171-14174.
66. F. Paesani, *Temperature-dependent infrared spectroscopy of water from a first-principles approach.* J. Phys. Chem. A, 2011. **115**(25): 6861-6871.
67. M. Paolantoni, N.F. Lago, M. Albertí, and A. Laganà, *Tetrahedral Ordering in Water: Raman Profiles and Their Temperature Dependence†.* J Phys Chem A, 2009. **113**(52): 15100-15105.
68. M. Smyth and J. Kohanoff, *Excess Electron Localization in Solvated DNA Bases.* Phys. Rev. Lett., 2011. **106**(23): 238108.
69. Y. Marechal, *Infrared-spectra of water. 1. Effect of temperature and of H/D isotopic dilusion.* J Chem Phys, 1991. **95**(8): 5565-5573.
70. G.E. Walrafen, *Raman Spectral Studies of the Effects of Temperature on Water Structure.* J Chem Phys, 1967. **47**(1): 114-126.
71. K. Furic and V. Volovsek, *Water ice at low temperatures and pressures: New Raman results.* J. Mol. Struct., 2010. **976**(1-3): 174-180.
72. H. Suzuki, Y. Matsuzaki, A. Muraoka, and M. Tachikawa, *Raman spectroscopy of optically levitated supercooled water droplet.* J Chem Phys, 2012. **136**(23): 234508.
73. Y. Marechal, *The molecular structure of liquid water delivered by absorption spectroscopy in the whole IR region completed with thermodynamics data.* J. Mol. Struct., 2011. **1004**(1-3): 146-155.
74. X.J. Liu, M.L. Bo, X. Zhang, L.T. Li, Y.G. Nie, H. TIan, S. Xu, Y. Wang, and C.Q. Sun, *Coordination-resolved bond-electron spectrometrics.* Chem. Rev., 2014. **Accepted 03/10/2014**.
75. T. Tokushima, Y. Harada, O. Takahashi, Y. Senba, H. Ohashi, L.G.M. Pettersson, A. Nilsson, and S. Shin, *High resolution X-ray emission spectroscopy of liquid water: The observation of two structural motifs.* Chem. Phys. Lett., 2008. **460**(4-6): 387-400.
76. J.H. Guo, Y. Luo, A. Augustsson, J.E. Rubensson, C. Såthe, H. Ågren, H. Siegbahn, and J. Nordgren, *X-Ray Emission Spectroscopy of Hydrogen Bonding and Electronic Structure of Liquid Water.* Phys. Rev. Lett., 2002. **89**(13): 137402.
77. A. Nilsson and L.G.M. Pettersson, *Perspective on the structure of liquid water.* Chem. Phys., 2011. **389**(1-3): 1-34.
78. J.J. Shephard, J.S.O. Evans, and C.G. Salzmann, *Structural Relaxation of Low-Density Amorphous Ice upon Thermal Annealing.* J. Phys. Chem. Lett., 2013: 3672-3676.
79. S. Iikubo, K. Kodama, K. Takenaka, H. Takagi, M. Takigawa, and S. Shamoto, *Local Lattice Distortion in the Giant Negative Thermal Expansion Material $Mn_3Cu_{1-x}Ge_xN$.* Phys. Rev. Lett., 2008. **101**(20): 205901.
80. A.L. Goodwin, M. Calleja, M.J. Conterio, M.T. Dove, J.S.O. Evans, D.A. Keen, L. Peters, and M.G. Tucker, *Colossal positive and negative thermal expansion in the framework material $Ag_3 Co(CN)(6)$.* Science, 2008. **319**(5864): 794-797.
81. A.C. McLaughlin, F. Sher, and J.P. Attfield, *Negative lattice expansion from the superconductivity-antiferromagnetism crossover in ruthenium copper oxides.* Nature, 2005. **436**(7052): 829-832.
82. J.S.O. Evans, *Negative thermal expansion materials.* J. Chem. Soc.-Dalton Trans., 1999(19): 3317-3326.
83. T.A. Mary, J.S.O. Evans, T. Vogt, and A.W. Sleight, *Negative thermal expansion from 0.3 to 1050 Kelvin in $ZrW_2O_8$.* Science, 1996. **272**(5258): 90-92.
84. S. Stoupin and Y.V. Shvyd'ko, *Thermal Expansion of Diamond at Low Temperatures.* Phys. Rev. Lett., 2010. **104**(8): 085901.
85. Q.H. Tang, T.C. Wang, B.S. Shang, and F. Liu, *Thermodynamic Properties and Constitutive Relations of Crystals at Finite Temperature.* Sci China-Phys Mech Astron, 2012. **G 55**: 933.
86. Y.J. Su, H. Wei, R.G. Gao, Z. Yang, J. Zhang, Z.H. Zhong, and Y.F. Zhang, *Exceptional negative thermal expansion and viscoelastic properties of graphene oxide paper.* Carbon, 2012. **50**(8): 2804-2809.
87. A.W. Sleight, *Compounds that contract on heating.* Inorg. Chem., 1998. **37**(12): 2854-2860.
88. J.S.O. Evans, T.A. Mary, T. Vogt, M.A. Subramanian, and A.W. Sleight, *Negative thermal expansion in $ZrW_2O_8$ and $HfW_2O_8$.* Chem. Mater., 1996. **8**(12): 2809-2823.

89. G. Ernst, C. Broholm, G.R. Kowach, and A.P. Ramirez, *Phonon density of states and negative thermal expansion in ZrW2O8.* Nature, 1998. **396**(6707): 147-149.
90. A.K.A. Pryde, K.D. Hammonds, M.T. Dove, V. Heine, J.D. Gale, and M.C. Warren, *Origin of the negative thermal expansion in $ZrW_2O_8$ and $ZrV_2O_7$.* J. Phys.-Condes. Matter, 1996. **8**(50): 10973-10982.
91. M. Chaplin. *Water structure and science: http://www.lsbu.ac.uk/water/*.